\documentstyle[newarcrc,fleqn,psfig]{article}

\def\plotfig#1#2#3#4#5#6#7{\centering \leavevmode
    \vbox to#2{\rule{0pt}{#2}}
    \includegraphics{#1}}

\title { Changes in the structure of the accretion disc \\
		of HS1804+67 through the outburst cycle \thanks{This work was
		partially supported by CNPq research grant no.\ 300\,354/96-7
		and by PRONEX grant FAURGS/FINEP 7697.1003.00.}}
\author{ Raymundo Baptista \address{Departamento de F\'\i sica - CFM, UFSC,
		88040-900 Florian\'opolis, SC, Brazil, email: bap@fsc.ufsc.br} and
		M. S. Catal\'an \address{Department of Physics, Keele University,
		Keele, Staffordshire, ST5 5BG, UK, email: msc@astro.keele.ac.uk}}

\begin{document}
\maketitle

\begin{abstract}
We report on the analysis of high-speed photometry of the dwarf-nova
HS1804+67 through its outburst cycle with eclipse mapping techniques.
Eclipse maps show evidences of the formation of a spiral structure in the
disc at the early stages of the outburst and reveal how the disc expands
during the rise until its fills most of the primary Roche lobe at maximum
light. During the decline phase, the disc becomes progressively fainter
as the cooling front moves inwards from the outer regions, until only a
small bright region around the white dwarf is left at minimum light.  
The variable part of the uneclipsed light is possibly due to emission in
a wind emanating from the inner parts of the disc. The emission from this
region is sensitive to the mass accretion rate.
\end{abstract}

\section {Introduction}

Dwarf novae yield an uneven opportunity to study the time evolution of
non-stationary accretion discs, in particular in the transition between
the high and low viscosity (and accretion rate) states which are believed
to occur during the outbursts of these objects. HS1804+67 is a long
period ($\simeq 5.5$ hs) eclipsing dwarf nova with outbursts of moderate
amplitude ($\simeq 1-2$ mag) and recurrence intervals of $\sim 1$ month
\cite{BMD95,FBM97}.

In this paper we report on the results of the analysis with eclipse mapping
techniques \cite{Horne} of a set of lightcurves of HS1804+67, which allows
to follow the evolution of the structure of its accretion disc through the
outburst cycle. The eclipse maps capture ``snapshots'' of the disc brightness
distribution on the rise to maximum, during maximum light, through the
decline phase, and at the end of the eruption -- when the system goes through
a phase of minimum light before recovering its quiescent brightness level.

\section {Observations and analysis}

Time series of CCD differential photometry of HS1804+67 in the R band
were obtained with the JGT 1-m telescope at the University of St.\,Andrews
during 1995-96 covering 4 consecutive eruptions of the star. The data were
grouped and average lightcurves were obtained for 11 different phases 
through the outburst cycle. The average lightcurves were analyzed with
eclipse mapping techniques to produce a map of the disc brightness
distribution and an additional uneclipsed component in each case.
The data lightcurves and corresponding eclipse mapping models are shown
in Fig.\,1. Maps of the brightness distribution for 9 of the lightcurves
are shown in Fig.\,2 in a logarithmic grayscale. Fig.\,3 shows the evolution
of the radial intensity distribution in the disc of HS1804+67 through the
outburst cycle.
%
%
\begin{figure}
\plotfig{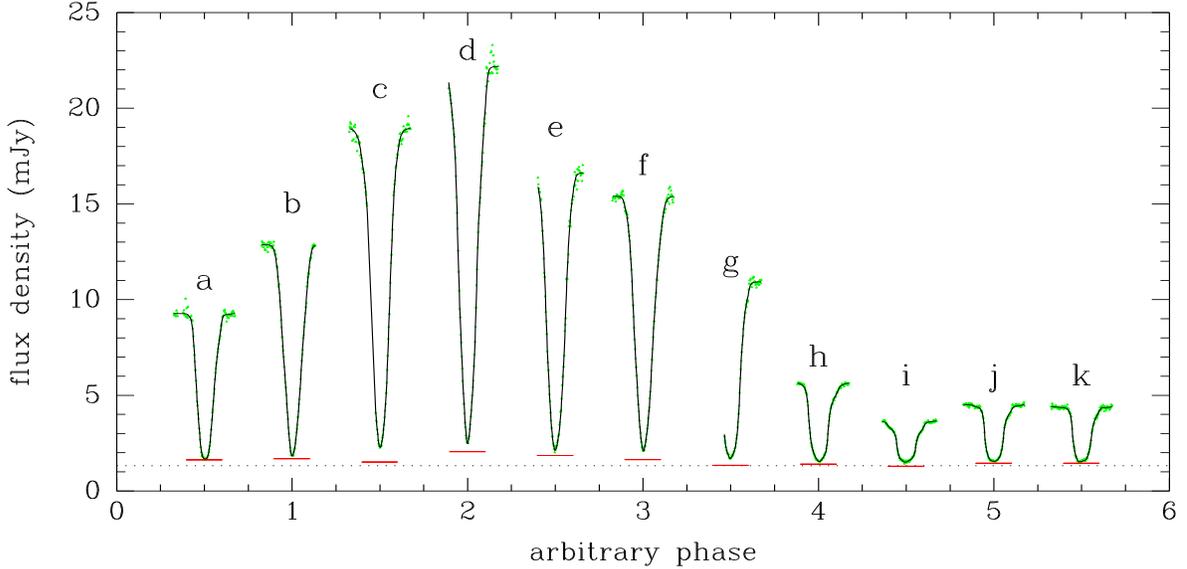}{6.5cm}{-90}{65}{65}{-255}{280}
\caption { Average lightcurves of HS1804+67 seen in a sequence through the
	outburst cycle. The separation of the lightcurves in the x axis is
	arbitrary. The solid lines are the eclipse mapping model lightcurves.
	The uneclipsed component in each case is indicated by a horizontal tick;
	a dotted horizontal line mark the value of the uneclipsed component at
	minimum light. }
\end{figure}

\section {Discussion}

The results reveal the formation of a spiral structure at the early
stages of the outburst (fig.\,2a) and shows how the disc expands until it
fills almost all of the primary Roche lobe at maximum light (figs.\,2c
and 3), becoming progressively fainter through the decline while the bright
spot starts to become more and more perceptible at the outer edge of the
disc (figs.\,2d-h and 3). 
At the phase of minimum the disc mostly disappears, leaving only a small
bright region around the white dwarf, possibly a boundary layer (fig.\,2i). 
In quiescence the disc is asymmetric, with the region along the gas stream
trajectory being noticeably brighter than the neighbouring regions
(fig.\,2j). This is is agreement with the results obtained from Doppler
tomography of H$\alpha$ emission \cite{BMD95}.
%
%
\begin{figure}
\plotfig{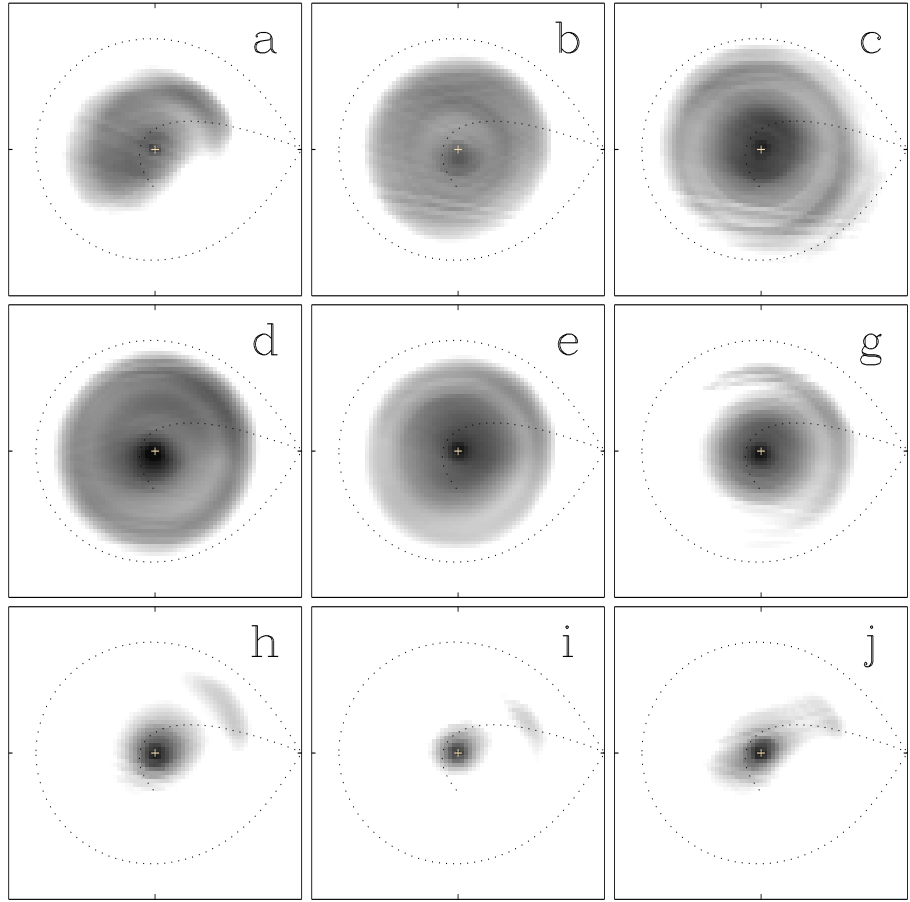}{8cm}{0}{70}{70}{-220}{-120}
\caption { Sequence of eclipse maps of HS1804+67 through the outburst cycle.
	Labels are the same as in Fig.\,1. Brighter regions are dark ($\log I=
	-2.8$) and fainter regions are white ($\log I= -5.6$). A cross mark the
	center of the disc; dashed lines shows the projection of the Roche lobe
	on the orbital plane and the gas stream trajectory; the secondary is to
	the right of each panel and the stars rotate counter-clockwise.}
\end{figure}
%
%
\begin{figure}[th]
\plotfig{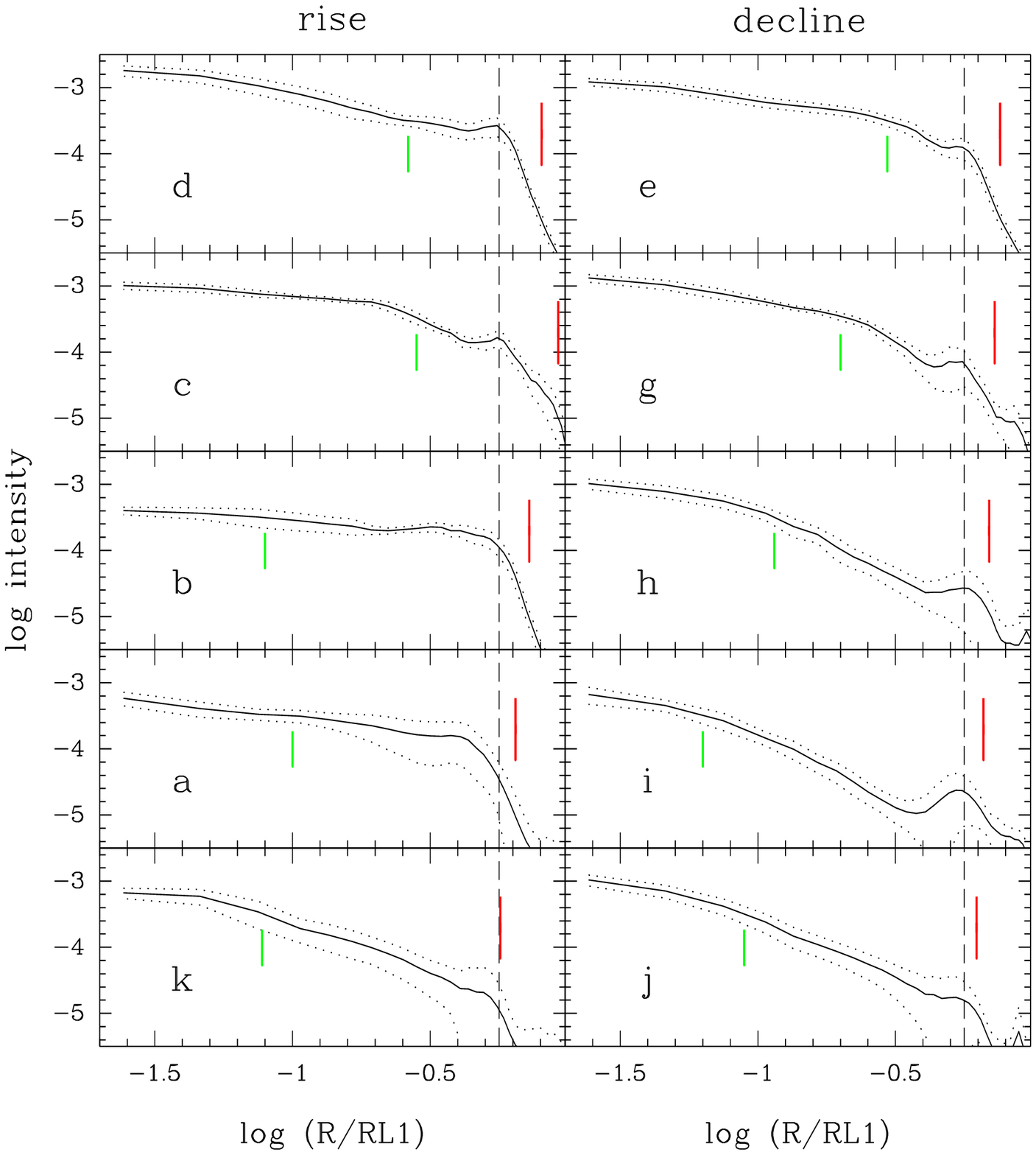}{10cm}{0}{70}{58}{-220}{-110}
\caption { The evolution of the radial intensity distribution through the
	outburst. Labels are the same as in Figs.\,1-2. R$_{L1}$ is the distance
	from disc centre to the inner Lagrangian point. Dashed lines indicate
	the radial position of the bright spot in quiescence. Vertical long ticks
	mark the position of the outer disc edge and vertical short ticks
	indicate the radial position at which the disc intensity drops below
	$\log I= -3.5$. }
\end{figure}

The comparison of the maps during the rise to maximum indicates that the
eruption starts in the outer disc and that a heating front wave (which
triggers the high viscosity and mass accretion state) moves inwards and
reaches the central parts of the disc at outburst maximum.
A cooling front wave characterizes the decline of the eruption and
also propagates from the outer parts to disc centre (fig.\,3) reaching
the central parts of the disc by the end of the outburst (fig.\,2h).
The comparison of the maps at minimum light and at quiescence suggests
that mass accretion over the white dwarf is substantially reduced in
the former phase and that probably most of the matter transferred from the
secondary star at these stages accumulates in the outer disc, restarting
the eruption cycle.

The evolution of the uneclipsed flux through the outburst can be seen in
Fig.\,1. A dotted line indicate the value of the uneclipsed component at
minimum light and is interpreted as being due to the (fixed) contribution
of the secondary star to the flux in the R band. The variable part of the
uneclipsed component is probably due to emission in a vertically extended
disc chromosphere + wind. Fig.\,1 shows that the emission from this latter
region follows the changes in brightness of the inner parts of the disc
during the outburst. At minimum light, when mass accretion at the inner
disc is substantially reduced, the emission in the disc chromosphere
+ wind practically disappears. These results support the suggestion that
the ejection of material in the wind originates from the inner parts of the
disc and that the emission of the resulting chromosphere + wind is
sensitive to the disc mass accretion rate, in accordance to inferences
drawn by a similar study of the novalike UX UMa \cite{B98}.


\begin{thebibliography}{ Billington etal 1995 }

\bibitem [Baptista et al, 1998]{B98}
Baptista R., Horne K., Wade R.A., Hubeny I., Long K., Rutten R.G.M.,
	1998, MNRAS, 298, 1079

\bibitem [Billington et al, 1995]{BMD95}
Billington I., Marsh T.R., Dhillon V.S., 1995, MNRAS, 273, 100

\bibitem [Fiedler et al, 1997]{FBM97}
Fiedler H., Barwig H., Mantel K-H., 1997, A\&A, 327, 173

\bibitem [Horne, 1985]{Horne} Horne, K. 1985, MNRAS, 213, 129

\end{thebibliography}
\end{document}